\documentclass[preprint,showpacs,preprintnumbers,amsmath,amssymb]{revtex4}

\usepackage{graphicx}
\usepackage{dcolumn}
\usepackage{bm}

\begin{document}


\title{Very Light Cosmological Scalar Fields from a Tiny Cosmological Constant}

\author{Xavier Calmet}
 \email{xavier.calmet@ulb.ac.be}
 \affiliation{Universit\'e Libre de Bruxelles, 
Service de Physique Th\'eorique, CP225,
Boulevard du Triomphe  (Campus plaine),
B-1050 Brussels, Belgium. 
}%

\date{\today}

\begin{abstract}
We discuss a mechanism which generates a mass term for a scalar field in an expanding universe. The mass of this field turns out to be generated by the cosmological constant and can be naturally small if protected by a conformal symmetry which is however broken in the gravitational sector. The mass is comparable today to the Hubble time. This scalar field could thus impact our universe today and for example be at the origin of a time variation of the couplings and masses of the parameters of the standard model.
\end{abstract}

\maketitle

Scalar fields have become  an important tool to build  cosmological models. They are used to explain a plethora of phenomena that range from inflation \cite{Lyth:1998xn} and dark energy \cite{Copeland:2006wr} to a time variation of the couplings and masses of the standard model particles \cite{Uzan:2002vq,Dvali:2001dd}.
A time dependent scalar field in an expanding universe can only fluctuate significantly  at a given time  in the evolution of  the universe if and only if its mass $m$ is comparable to the Hubble time $H$ (to be precise $m>3/2H$)
\begin{eqnarray} \label{scalar}
\ddot \phi + 3 H \dot \phi + m^2 \phi +...=0.
\end{eqnarray}
This equation can be derived easily from the expansion of a scalar field in an expanding universe using the Robertson-Walker metric. It is easy to see that indeed if that mass of the field is much smaller than $H$, the friction term dominates and the field does not oscillate, however if the mass is much bigger than the present Hubble time, the field would start to oscillate much earlier in the history of the universe and would have reached a minimum long ago and would not affect the present universe.

Most cosmological models are trying to address early cosmological phenomena such as inflation and it is not difficult to imagine mechanisms to explain a large scalar mass of the order of the Hubble time. A simple usual mass term $m^2 \phi \phi$ in the action describing the dynamics of the scalar field would do the trick. However, if the scalar field is to impact the universe today, i.e. $H\sim 10^{-33}$eV, one needs to explain a very tiny scalar mass  \cite{Dvali:2001dd}.

This is what is required to generate a time variation of physical parameters today. There are different astrophysical observations  \cite{Webb:1998cq,Webb:2000mn,Murphy:2003hw,Murphy:2006vs,Reinhold:2006zn,Ivanchik:2002me} indicating  a cosmological evolution of the gauge couplings and masses of the particles of the standard model. On the other hand, other groups \cite{Chand:2004ct,Srianand:2004mq,Chand:2006va} however have not observed this effect.  This is clearly an observational issue which needs to be settled down by astronomers. From a theorist point of view the main difficulty is to understand how a scalar field  with a mass of the order of $m\sim 10^{-33}$eV could be generated. The main goal of this work is to propose a new mechanism to generate a very light scalar mass in an expanding universe.

We shall consider a scalar field which is not minimally coupled to gravity and focus on the operator \cite{Feynman:1996kb}
\begin{eqnarray} \label{newop}
\alpha \int d^4x \sqrt{-g} R \phi \phi,
\end{eqnarray}
where $\alpha$ is a dimensionless coupling constant, $R$ is the curvature scalar and $\phi$ is a scalar field. This action is invariant under general coordinate transformations, invariant under Lorentz transformations  and it would be gauge invariant if the scalar field was a gauge field. We use the signature $(+1,-1,-1,-1)$. The full gravitational action is then 
\begin{eqnarray} \label{action}
  \int d^4x \sqrt{-g}  \left ( \frac{1}{16 \pi G}(R -2 \Lambda) + \frac{1}{2} g^{\mu\nu}\partial_\mu\phi \partial_\nu\phi  - \frac{\alpha}{2} R \phi^2\right)
\end{eqnarray}
where $G$ is the gravitational coupling and $\Lambda$ is the cosmological constant. We note that  a very similar action has been studied already in the literature see e.g. \cite{Wetterich:1987fk,Wetterich:1987fm,Wetterich:1994bg,Wetterich:2002wm,Fujii:2002sb} and is closely related to the Jordan-Brans-Dicke \cite{Jordan,Brans:1961sx} theory of gravity.  The main difference in our approach is that we have included a term $ \sqrt{-g}  \frac{1}{16 \pi G}R$ and a cosmological constant explicitly. These differences have important consequences. In particular our model contains a conformal violating term, i.e. the usual Hilbert-Einstein term and the action thus cannot be mapped into that studied in  \cite{Wetterich:1987fk,Wetterich:1987fm,Wetterich:1994bg,Wetterich:2002wm,Fujii:2002sb}. Furthermore, our scalar field is not the main component of the dark energy like  the cosmon in the scenario proposed by Wetterich \cite{Wetterich:1987fk,Wetterich:1987fm,Wetterich:1994bg,Wetterich:2002wm}. Our approach is different, we are not trying to find an alternative to a cosmological constant. However we want to study the consequences of the observed cosmological constant for a scalar field coupled in a non-minimal way to gravity. 

 The field equations obtained from the action (\ref{action}) are given  by
\begin{eqnarray}
  R_{\mu\nu} - \frac{1}{2} g_{\mu\nu} R +  \Lambda g_{\mu\nu} =  - 8 \pi G S_{\mu\nu}
  \end{eqnarray}
  with
  \begin{eqnarray}
 S_{\mu\nu} = 
   \left ( 
    \partial_\mu\phi \partial_\nu \phi - \frac{1}{2} g_{\mu\nu} \partial_\rho \phi  \partial^\rho \phi - \alpha \phi^2 R_{\mu\nu}+\frac{1}{2} g_{\mu\nu} \alpha  \phi^2 R-  \alpha (g_{\mu\nu} g^{\alpha\beta} \phi^2_{;\alpha\beta} - \phi^2_{;\mu\nu})  \right)
  \end{eqnarray}
 and
  \begin{eqnarray}
 g^{\mu\nu} \nabla_\mu \nabla_\nu \phi + \alpha R \phi =0
   \end{eqnarray}
   where $\alpha R$ plays the role of a mass term for the scalar field. 
   
To study the connection to Brans-Dicke type models, it is useful to rewrite Einstein's equations as    
   \begin{eqnarray}
  R_{\mu\nu} - \frac{1}{2} g_{\mu\nu} R +  \frac{G_{eff}}{G} \Lambda g_{\mu\nu} = -8 \pi G_{eff} \left ( \partial_\mu\phi \partial_\nu\phi -\frac{1}{2} g_{\mu\nu} \partial_\rho \phi  \partial^\rho \phi -  \alpha (g_{\mu\nu} g^{\alpha\beta} \phi^2_{;\alpha\beta} - \phi^2_{;\mu\nu})   \right )
  \end{eqnarray}
where we have introduced an  effective gravitational coupling constant 
 \begin{eqnarray} \label{effGN}
G_{eff}= \frac{G}{1-8\pi G \alpha \phi^2}
   \end{eqnarray}
   Note that the effective gravitational coupling constant  $G_{eff}$ is space-time dependent and could easily lead to a space-time dependency of the parameters of the standard model because of quantum effects and in particular through the renormalization group equations as discussed in \cite{Calmet:2001nu,Calmet:2002ja,Calmet:2002jz,Calmet:2006sc}. A variable Newton constant has first been discussed in \cite{Zee:1978wi}, see also \cite{Dirac,Wu:1986ac,Wetterich:1987fk}. Furthermore, if one adopts an effective theory point of view as in e.g. \cite{Dvali:2001dd}, it is natural to couple the scalar field $\phi$ to the fields of the standard model which will lead to a fifth force type new interaction and would be another source of time variation for the parameters of the standard model.

  Let us now go back to our consideration on the equations of motion and make use of the contracted Einstein equations 
  \begin{eqnarray}
  -R+4\Lambda=  8 \pi G( \partial_\mu \phi \partial^\mu \phi - \alpha \phi^2 R + 3 \alpha \nabla_\alpha \nabla^\alpha \phi^2)
  \end{eqnarray}
  and we obtain
 \begin{eqnarray}
 g^{\mu\nu} \nabla_\mu \nabla_\nu \phi + 4 \alpha \Lambda \phi  - 8 \pi G \alpha \phi (\partial_\mu \phi \partial^\mu \phi - \alpha \phi^2 R + 3 \alpha \nabla_\mu \nabla^\mu \phi^2)=0.
   \end{eqnarray}
  We see that the scalar field acquires a mass term given by $m=2 \sqrt{\alpha \Lambda}$ if there is a cosmological constant $\Lambda$. Using $\Lambda= 8\pi G \rho_{vac}$ and the measured vacuum density \cite{Perlmutter:1998np} i.e. $ \rho_{vac}\sim (2.4 \times 10^{-3}$ eV$)^4$, we get $m=4.7 \times 10^{-33}$ eV assuming that $\alpha$ is of order one. This is the right order of magnitude. If $\phi$ is only time dependent and given the measured energy density, a time variation of the physical parameters is therefore not surprising. 
  
  It is easy to find a solution to equation (\ref{scalar}) which describes a scalar field in an expanding universe. One finds
  \begin{eqnarray}
  \phi(t) = \mbox{Re}(c_1 \exp{(w_1 t)} + c_2  \exp{(w_2 t)})
  \end{eqnarray}
   with
   $w_{1/2}= - 3/2 H \pm  \sqrt{9/4 H^2 -m^2}$. In order to have an oscillatory behavior, $w_{1/2}$ has to be complex which is possible if $m>3/2H$ which is, as explained above, possible today in the universe if $m\sim 3/2 H$. If the mass was much smaller than the Hubble scale, there would be no oscillation and and if it was much larger than today's value of $H$, the oscillation would have taken place at an earlier time in the evolution of the universe and the scalar field would not impact today's universe.

 We can now estimate the time change in the Newton gravitation coupling constant between its value at the Big Bang and today's value assuming that oscillations take places today (i.e. $m\sim 3/2 H_0$). We find $(G_{eff}(t_0)-G_{eff}(0))/G_{eff}(t_0)=-8 \pi G \alpha \Delta \phi^2/(1-8 \pi G \alpha \phi^2)$ where $t_0$ is the age of the universe today. For the Planck scale $\Lambda_{Planck}=1/\sqrt{G}$ this implies
$\Delta \Lambda^2_{Planck}= -8 \pi \alpha \Delta \phi^2$. If we take  $\Delta \phi^2$ of the order of the Planck scale (i.e. $(c_1+c_2)^2 \sim \Lambda^2_{Planck}$),
 this can be a sizable effect and it  is not difficult to imagine that the  observations of  Webb et al. \cite{Webb:1998cq,Webb:2000mn,Murphy:2003hw,Murphy:2006vs,Reinhold:2006zn,Ivanchik:2002me} could be explained by a renormalization group effect and because of the time dependence of the Planck scale. Such an effect is characteristic of a theory which unifies gauge interactions and gravity such as Kaluza-Klein theories, see e.g. \cite{Marciano:1983wy}.  Note that the time change of the Planck scale is given by
 \begin{eqnarray}
 \frac{\dot \Lambda_{Planck}(t)}{\Lambda_{Planck}(t)}= -8 \pi \alpha \frac{\dot \phi(t) \phi(t)}{\Lambda^2_{Planck}(t)}.
 \end{eqnarray}
 Obviously a time variation of physical constants is only observable if some other scale remains constant or a least changes with time at a different rate. This could be for example the scale where fermion masses are generated, scale which is not obviously related to the Planck scale.

  As mentioned previously the action (\ref{action}) is very similar to a Jordan-Brans-Dicke theory defined by
  \begin{eqnarray}
    \int d^4x \sqrt{- g}  \frac{1}{16 \pi }\left (\Phi  R  -  \omega \frac{  g^{\mu\nu}\partial_\mu\Phi \partial_\nu\Phi}{\Phi} \right ),
  \end{eqnarray}
  where we have omitted the cosmological constant.
 However, there are serious constraints on this theory of gravitation. The coefficient of this theory, $\omega$ has to be greater than 500 to avoid conflicts with observations \cite{Reasenberg}. Our action can be mapped to that  of Jordan-Brans-Dicke. The Jordan-Brans-Dicke parameter is then given by $\omega= (1-8  \pi \phi^2 G \alpha)/(32 \pi \phi^2 G \alpha^2)$ and is dependent on the scalar field, however let us  assume that $\phi$ oscillates slowly over time lapses relevant for gravitational measurements in the solar system and thus consider $\omega$ to be constant. The bound on $\omega$ implies the bound $\phi/\Lambda_{Planck} < 4 \times 10^{-3}$  assuming $\alpha \sim 1$ and there is thus much space for a time variation as discussed above. We note that strictly speaking we are not dealing with a Jordan-Brans-Dicke theory since our $\omega$ is not constant however similar bounds do apply because the trace of the energy momentum tensor is a source for our scalar field as well.
  It is interesting to note that the bound on $\omega$ implies a bound on the time change of the Newton coupling constant since the Big Bang which is of the order of $|\Delta G/G| < 4 \times 10^{-4}$.

  Let us now study the couplings of the new scalar sector to the standard model. The only dimension four operator is the four scalar coupling $h^\dagger h \phi^2$ where $h$ is the Higgs field of the standard model. This coupling is dangerous since, after the Higgs field has acquired a vacuum expectation value, it would lead to a new contribution to the mass of the scalar field $v^2 \phi^2$ with $v=246$ GeV which would clearly dominate the mass of the cosmological scalar field. It is thus tempting to assume that cosmological scalar field couples as a dilaton to matter. As emphasized in \cite{Dvali:2001dd} (see also \cite{Chiba:2006xx}), this would lead to the desired time variation of the hyperfine structure constant.  This is quite natural in the framework of a string theory dilaton \cite{Damour:1994zq}. However, a dilaton type coupling seems to generically generate the operator  $h^\dagger h \phi^2$ at the quantum level. The strength of this operator is model dependent, this operator, however, seems difficult to avoid. This is a strong indication that the cosmological scalar field under consideration can only couple gravitationally to the standard model, in which case its mass is naturally small. As explained above its main effect is then to  impact the strength of the  gravitational coupling constant and hence the Planck scale.

  Let us now discuss the cosmology of our model. The action considered in  equation (\ref{action}) can be rewritten in the Einstein frame using   the transformations \cite{Schmidt:1988xi,Dick:1998ke}:
  \begin{eqnarray}
  \phi= \sqrt{\frac{1}{\alpha}} M_{r} \tanh \left (\frac{\hat \phi \sqrt{\alpha}}{ M_r} \right)
  \end{eqnarray}
   and
    \begin{eqnarray}
g_{\mu\nu}= \cosh^2\left (\frac{\hat \phi \sqrt{\alpha}}{M_r} \right) \hat g_{\mu\nu}
  \end{eqnarray}
  where $M_r=\sqrt{1/(8\pi G)}$ is the reduced Planck mass.
  \begin{eqnarray} \label{actionEinstein}
  \int d^4x \sqrt{-\hat g}  \left ( \frac{1}{16 \pi G}\left (\hat R -2 \Lambda \cosh^4\left (\frac{\hat \phi \sqrt{\alpha}}{ M_r} \right) \right ) + \frac{1}{2} \hat g^{\mu\nu}\partial_\mu \hat \phi \partial_\nu \hat  \phi \right).
\end{eqnarray}
We obtain a scalar field coupled minimally  to gravity. If we look into more details into the vacuum energy term we recover our previous result. If we assume that $\phi<<M_r$, which has to be the case to avoid conflicts with the bounds on the Jordan-Brans-Dicke parameter, we can expand the $\cosh$ term and obtain 
   \begin{eqnarray} \label{actionEinstein2}
  \int d^4x \sqrt{-\hat g}  \left ( \frac{1}{16 \pi G}\left (\hat R -2 \Lambda \right ) -2 \alpha \Lambda \hat \phi^2 -\frac{5}{24\pi G} \alpha^2 \hat  \Lambda \phi^4 - {\cal O}\left(\frac{\hat \phi}{M_r}\right)^6  + \frac{1}{2} \hat g^{\mu\nu}\partial_\mu \hat \phi \partial_\nu \hat  \phi \right).
\end{eqnarray}
  i.e. we obtain a massive scalar field  with a $\phi^4$ term coupling in a minimal manner to gravity. An important point can be made at this point. The cosmological constant of our universe is positive, if this scalar field was gauged it would have a positive squared mass (if $\alpha >0$ which, as we will see, is the case if the scalar field is coupled in a conformal manner to gravity) and it would  hence not lead to the Higgs mechanism. In a anti-de-Sitter world, the mass squared would be negative  if $\alpha >0$, however the $\phi^4$ term would be negative as well and this independently of the sign of $\alpha$. It thus seems that to induce spontaneous gauge symmetry breaking through a gravitational interaction one needs to have a de Sitter cosmological constant and a negative parameter $\alpha$. This discussion would be obviously changed if we had included a term $\lambda \phi^4$ from the beginning since we would have introduced a free parameter $\lambda$ by hand and spontaneous gauge symmetry breaking would be possible both in the de Sitter and anti-de-Sitter cases depending on the values of $\alpha$ and $\lambda$.

Up to this point we had to rely on fine-tuning to explain a small scalar mass.   We can now refine the argument and try to address the question of why is  $\alpha$ of order one. It turns out that a symmetry, local conformal invariance, which has been considered in \cite{Deser:1970hs,Callan:1970ze} and is described in e.g. \cite{Birrell:1982ix} fixes the parameter $\alpha$ to be equal to $1/6$ and furthermore prohibits a mass term of the type $m^2 \phi^2$ for the scalar field.  
  
  We note that the idea that scale invariance might be relevant to theories with scalar fields and in particular to  the standard model is not new. Indeed as pointed out by Bardeen \cite{Bardeen:1995kv}, if one sets the Higgs boson's mass to zero in the standard model, the model has no intrinsic scale at tree level. Obviously, the Higgs boson would get a mass through the Coleman-Weinberg mechanism \cite{Coleman:1973jx,Gildener:1976ih}. This idea has led to diverse models see e.g. \cite{Meissner:2006zh,Foot:2007as} for some recent papers. It is conceivable that the standard model and gravity could emerge out of a conformal theory. It is worth pointing out that  classical gravity, as long as it is not coupled to matter, is scaleless. In the same sense, one can show that the Planck length appears only when quantum mechanics and general relativity are considered together. This can be  shown by considering thoughts experiments which are designed to measure the shortest possible length \cite{minlength1,Calmet:2004mp}. One might thus speculate that nature should be described by an action which is  scaleless at the classical level, but that all dimensionfull parameters are generated by quantum effects.  We will assume that conformal invariance remains exact in the cosmological scalar sector as long as there is no cosmological constant which will induce a tiny scale invariance breaking in that sector.
  
  Since we are dealing with rather exotic physics, it is conceivable that general relativity and especially the sector describing  dark energy  break down at some length scale in our framework and it is thus best to consider the conformal scalar field in an expanding universe without imposing the constraint that it is described by general relativity.
  The action for a field coupled to the metric in a locally conformal invariant way is given by 
  \begin{eqnarray}
  \int d^4x \frac{1}{2} \sqrt{-g} \left ( g^{\mu\nu} \partial_\mu \phi \partial_\nu \phi 
  -\frac{1}{6} R \phi^2 \right).
 \end{eqnarray}
It is possible to add a further term in the action $ \sqrt{-g} \lambda_\phi \phi^4$ which is invariant under conformal transformations as well. However, if the scalar field turned out to be gauged, this term would induce a mass term at the one loop level and would make the conformal invariance anomalous, we would then have to set $\lambda_\phi=0$. However this term is not important in our discussion.

   The expansion of the scalar field in a Robertson-Walker metric is given by
  \begin{eqnarray}
  \ddot \phi + 3 H \dot \phi + (1-q)  H^2 \phi=0,
 \end{eqnarray}
where $H=\dot a/a$ and $q$ is the deceleration parameter given by \cite{Copeland:2006wr}
\begin{eqnarray}
  q(z)= \frac{3}{2} \frac{\sum_i \Omega_i^{0} (1+\omega_i) (1+z)^{3(1+\omega_i)}}{\sum_i \Omega_i^{0} (1+z)^{3(1+\omega_i)}} -1
 \end{eqnarray}
where $z$ is the redshift, $\omega_i$ is the equation of state for the $i$-th form of energy present in the universe and $\Omega_i^0$ is the corresponding dimensionless density parameter.  We thus have a prediction for the evolution of the mass of the scalar field which is independent of general relativity and characteristic of the conformal symmetry. If the time variation of the coupling constant is due to a conformal field the effect will evolve with the time dependent scalar mass
\begin{eqnarray}
 m(z) = \sqrt{(1-q(z))} H(z)
  \end{eqnarray}
  which can be checked by cosmological observations independently of verifying general relativity. Using $\Omega_m^0=0.3$ and $\Omega_\Lambda=0.7$ for today's universe we find $q(0)=-0.55$ and hence
  $m(0)=1.9 \times 10^{-33}$ eV, in accordance with our previous estimate.  A time variation of physical parameters such as the fine structure constant  in recent cosmological time could thus be explained by a scalar field coupled in a conformal manner to gravity.  This should be a strong motivation to continue to search for time variation of physical parameters using atomic clock experiments \cite{Fischer:2004jt}.

 {\it Acknowledgments:} The author would like to thank  Johann Rafelski for numerous and enlightening discussions which led to this work.  He is also grateful to Stephen Hsu and Peter Tinyakov for helpful suggestions. Finally, he would like to thank an anonymous referee for valuable comments and questions which led to an improvement of this work. This work was supported in part by the IISN and the  Belgian science policy office (IAP V/27).

\end{document}